\documentclass[aps,prb,twocolumn,superscriptaddress,showpacs,nofootinbib,longbibliography]{revtex4-2}

\usepackage{amsmath,amssymb,bm}
\usepackage{graphicx,xcolor}
\usepackage[colorlinks=true,linkcolor=blue,citecolor=blue,urlcolor=blue]{hyperref}

\newcommand{\kk}{{\bm{k}}}

\newcommand{\tauv}[1]{{\hat{\tau}_{#1}}}

\renewcommand{\vec}[1]{\bm{#1}}

\begin{document}
	
	\title{Light-Cone Friedel Oscillations: a Window onto van Hove Superconductivity}
	
	\author{Archisman Panigrahi}
	\affiliation{Department of Physics, Massachusetts Institute of Technology, Cambridge 02139, USA}

	\author{Vladislav Poliakov}
	\affiliation{Department of Physics, Massachusetts Institute of Technology, Cambridge 02139, USA}
	
	\author{Leonid Levitov}
	\affiliation{Department of Physics, Massachusetts Institute of Technology, Cambridge 02139, USA}
	
	
	\begin{abstract}
Van Hove (vH) singularities can enhance superconductivity by generating a large pairing susceptibility. Yet, it remains challenging to experimentally establish whether pairing occurs near these points. We show that saddle-point dispersion near a vH point gives rise to a ‘light-cone’ vH-selective behavior of Friedel oscillations, which is drastically different from the familiar concentric circles. Each vH point maps onto a real-space light-cone sector, allowing STM around an impurity to measure whether—and how strongly—the saddle-point quasiparticle sector is gapped. A superconducting gap near vH points suppresses the oscillations in the directions corresponding to peaks in the gap function angular dependence, a behavior that provides a signature of vH-related pairing at STM-accessible length scales. This STM protocol, targeting vH superconductivity, can determine whether superconducting pairing emerges near the vH points or elsewhere on the Fermi surface.
	\end{abstract}
	
	\maketitle
	
	
Van Hove (vH) points~\cite{VanHove_the_1953}---saddle points of the electronic dispersion---can strongly influence correlated metals: when the Fermi level lies nearby, their enhanced density of states can strengthen 
pairing susceptibility and drive unconventional superconductivity~\cite{
Hirsch1986,Dzyaloshinskii1987a,Dzyaloshinskii1987,Markiewicz1997,Gonzalez2000,Nandkishore2012,Steppke_strong_2017,
Ojajarvi_pairing_2024,Gil2026}; 
It can also generate non-Fermi-liquid behavior~\cite{Gonzalez1996,GonzalezStauber2020,Chronister2022,Stangier2022,Hunter2025},
and competing charge or magnetic instabilities~\cite{Markiewicz1997,Furukawa1998,LeHurRice2009,Schulz1987}. Here we discuss vH-selective Friedel oscillations and show that they provide a 
unique window onto carrier dynamics and pairing near these singular regions of $k$ space. We will focus on vH-assisted superconductivity, arguably the most intriguing and widely studied behavior. 



\begin{figure}[t]
	\centering
	\includegraphics[width=0.99\linewidth]{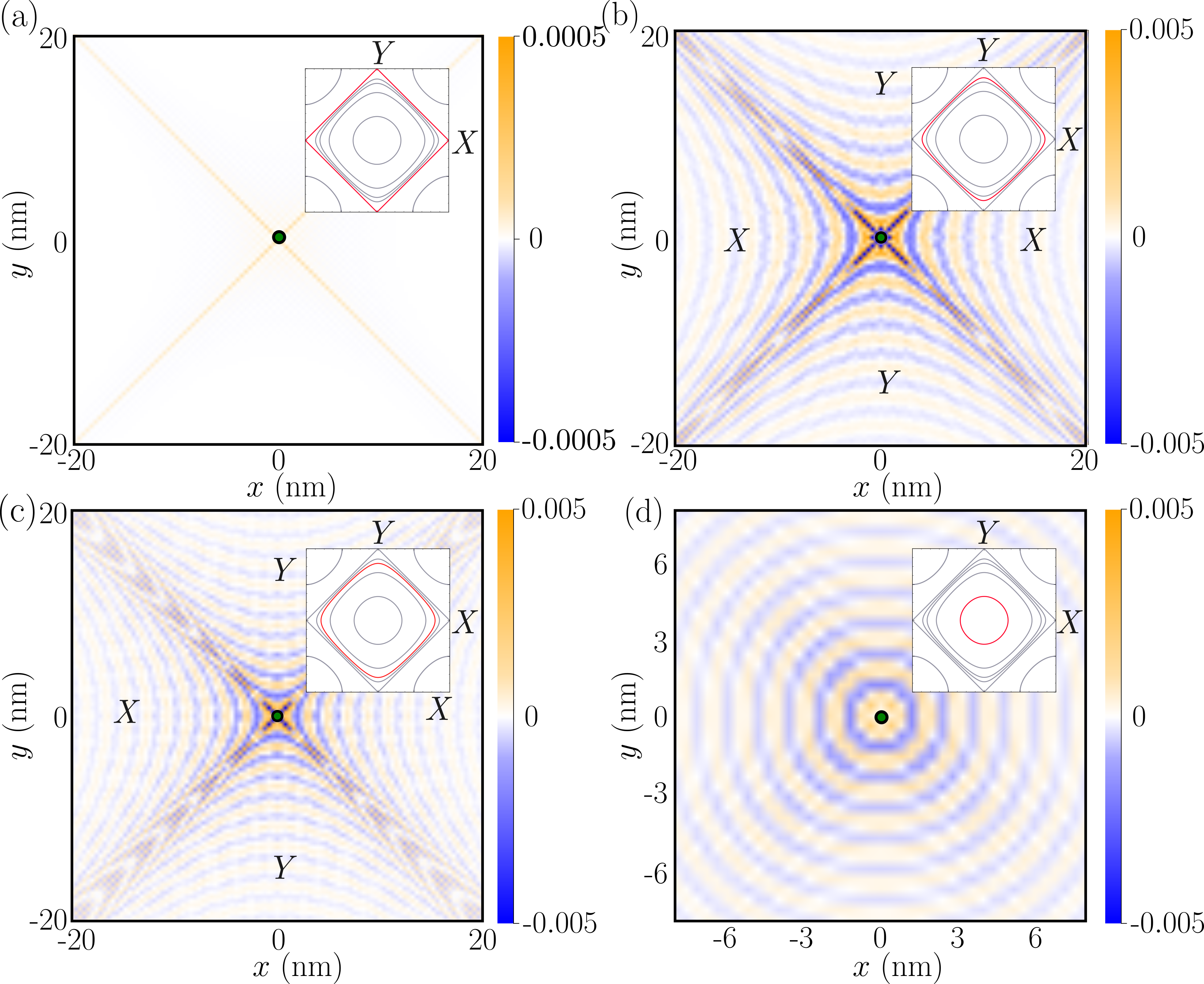} 
\caption{
(a–d) 
Light-cone Friedel oscillations around a point-like impurity in a square-lattice band. 
At small nonzero doping, the interference fringes form hyperbolic patterns [Eq.~\eqref{eq:hyperbola}], with spacing decreases as $1/\sqrt{|\mu|}$ with Fermi level detuning from the vH point increasing [panels (b) and (c)]. The oscillations are vH-selective: each vH point in momentum space generates hyperbolic fringes confined to a distinct, non-overlapping pair of real-space sectors ($X$ or $Y$); in each sector, the hyperbolas asymptotes to the sector boundaries. At half filling, $\mu=0$, the oscillatory fringes vanish [panel (a)], whereas at larger detuning they evolve into concentric circular fringes [panel (d)]. Shown is the tunneling density of states [Eq.~\eqref{eq:dIdV}], calculated for lattice constant $a=0.4$ nm and $\mu=0$, $-0.2t$, $-0.4t$, and $-3t$ in panels (a)–(d), respectively. 
Friedel oscillations arising from vH points are highly sensitive to superconductivity at these points (see Fig.~\ref{fig:Friedel_SC_nonSC}).
} 
\vspace{-2mm}
	\label{fig:vh}
\end{figure}

Theory predicts that vH-enhanced scattering phase space can produce a momentum-space hot-spot structure in which the superconducting gap is largest near the vH points
~\cite{Hirsch1986,Dzyaloshinskii1987a,Dzyaloshinskii1987,Markiewicz1997,Gonzalez2000,Nandkishore2012,
Ojajarvi_pairing_2024}. Yet, this is not a universal correspondence, as the gap structure may also depend on the pairing interaction and the symmetries of the Wannier functions~\cite{Choubey2014,Hirschfeld2015,Sharma2021,Du2018}. 
Whether superconducting pairing is in fact concentrated in the vH regions is therefore material dependent and must be determined experimentally~\cite{Gofron1994,Piriou2011,Levy2008,Newns_quasiclassical_1994}. Resolving this question requires a probe that can distinguish superconductivity at vH points. Our goal is to provide a practical real-space probe complementary to ARPES, that directly links these two phenomena, by determining whether the quasiparticles associated with individual vH points become gapped by the superconducting order.

On the experimental side, 
there are several actively-studied STM-able superconducting systems where van Hove physics is believed to be important, such as  cuprates~\cite{Markiewicz2023}, magic-angle twisted bilayer graphene and related moir'e systems~\cite{Bistritzer2020,Cea2021}, strain-tuned $~\mathrm{Sr_2RuO_4}$~\cite{Mueller2024,Stangier2022}, and kagome superconductors such as $\mathrm{CsV_3Sb_5}$~\cite{Wu2021,Xu2023}.
Yet, it is not known whether the vH singularity acts merely as a strong amplifier of low-energy correlations and pairing tendencies, or as a universal microscopic mechanism.

With this motivation, here we propose a scanning-tunneling-microscopy (STM) protocol capable of determining whether superconducting pairing is concentrated near the van Hove saddle points, or away from them. Near these points, impurities generate a characteristic vH-selective ‘light-cone’ Friedel oscillations that depend on Minkowski distance, with constant-phase contours
\begin{equation}\label{eq:hyperbola}
		s = \sqrt{(x-x_i)^2 - (y-y_i)^2} = n \frac{\lambda_F}2,\quad 
		\lambda_F = \frac{2 \pi}{k_F},
\end{equation} 
where $(x_i,y_i)$ are the impurity coordinates. 
Hyperbolic interference fringes have arisen previously in several distinct contexts~\cite{Aristov1997,Houghton2019,Bena2016,Uldemolins_quasiparticle_2022,Ortuzar_yu-shiba-rusinov_2022}, but, to our knowledge, have not been developed as a real-space diagnostic of vH-selective superconducting pairing.


Oscillatory functions of $s$, replacing the conventional circular Friedel oscillations, occur in separate `light-cone' sectors, which form distinct pairs, one per each vH point. {In the normal state, at $T=0$, the oscillations display power-law decay in vH-specific sectors and rapid exponential decay in 
the complementary sectors. In a superconducting state, the power-law decay turns into an exponential decay [see Eq.\eqref{eq:light-cone-behavior}]. In Fig.\ref{fig:vh} this behavior is illustrated for the two-dimensional square lattice band, 
\begin{equation}\label{eq:square_lattice_bandstructure}
E_{\vec k}=-2t\cos(k_xa)-2t\cos(k_ya),
\end{equation}
where vH points occur at the $X$ and $Y$ symmetry points $(\pi,0)$ and $(0,\pi)$. 
	At half-filling, localized impurities produce the familiar beam-like features in the tunneling current~\cite{Salkola_theory_1996,Balatsky2006,Mahan1995,Andrade2010,
	Weismann2009,Lounis2011,Byers_influence_1993,Das_local_2011}. 
Slightly away from half-filling, analysis predicts spatial oscillations behaving as  
\begin{equation}\label{eq:light-cone-behavior}
	\frac{dI}{dV}(\vec r) \sim u_0\frac{\cos(2k_F s)}{s}e^{-2s/\xi}, 
	\quad 
	k_F=\frac{\sqrt{|\mu|}}{ta^2}
\end{equation}
where $\xi(\Delta,T)$ is the superconducting coherence length given in Eq.\eqref{eq:coherence-length}, 
$\mu$ is Fermi level detuning from vH point and $s$ is the Minkowski distance given in Eq.\eqref{eq:hyperbola}, taken in the semiclassical limit $k_Fs\gg1$. The constant phase contours 
(nodes of $\cos 2k_F s$) are hyperbolas with periodicity varying with the detuning $\mu$ from vH point, as $1/\sqrt{|\mu|}$. At larger $\mu$ the fringes transform into conventional concentric circles. We derive these interference patterns using a continuum saddle-point model with Bogoliubov-de Gennes Green functions and a T-matrix treatment of impurity scattering, and verify them numerically in a realistic tight-binding model on a square lattice.

Our analysis identifies observable signatures of vH points and superconductivity concentrated at these points accessible by state-of-the art STM measurements. For doping close to vH point, $|\mu|\ll t$, the periodicity of the light-cone Friedel oscillations is much greater than atomic spacing, varying with $\mu$ as $1/\sqrt{|\mu|}$. The larger-than-atomic periodicity and a characteristic dependence on detuning from vH point provide clear signatures of the presence of vH points in the  bandstructure. 
A measured suppression of saddle-point fringes in the superconducting state would provide evidence for a gap affecting saddle-point quasiparticles. The decay of light cone oscillations with distance, which is governed by the thermal coherence length $\xi(T)=\hbar v_F/\pi k_B T$ in the normal state changes to a shorter lengthscale 
\begin{equation}
\xi(\Delta,T)=\frac{\hbar v_F}{\sqrt{\Delta^2+\pi^2 k_B^2 T^2}}
\label{eq:coherence-length}
\end{equation}
in the superconducting state, a behavior that can be used to determine the value of $\Delta$. Since $v_F$ scales with vH detuning as $\sqrt{|\mu|}$, for  doping values $\mu$ close to vH point, the corresponding lengthscales can be in the tens of nanometers (see Fig.~\ref{fig:vh}, and also Figs. \ref{fig:Friedel_SC_nonSC} and \ref{fig:Friedel_nonSC_different_lattice} discussed later). This is much shorter than in typical superconducting metals with a large Fermi surface, falling in the range readily accessible by state-of-the-art STM probes. 

	\begin{figure}[t!]
	\centering
	\includegraphics[width=0.99\linewidth]{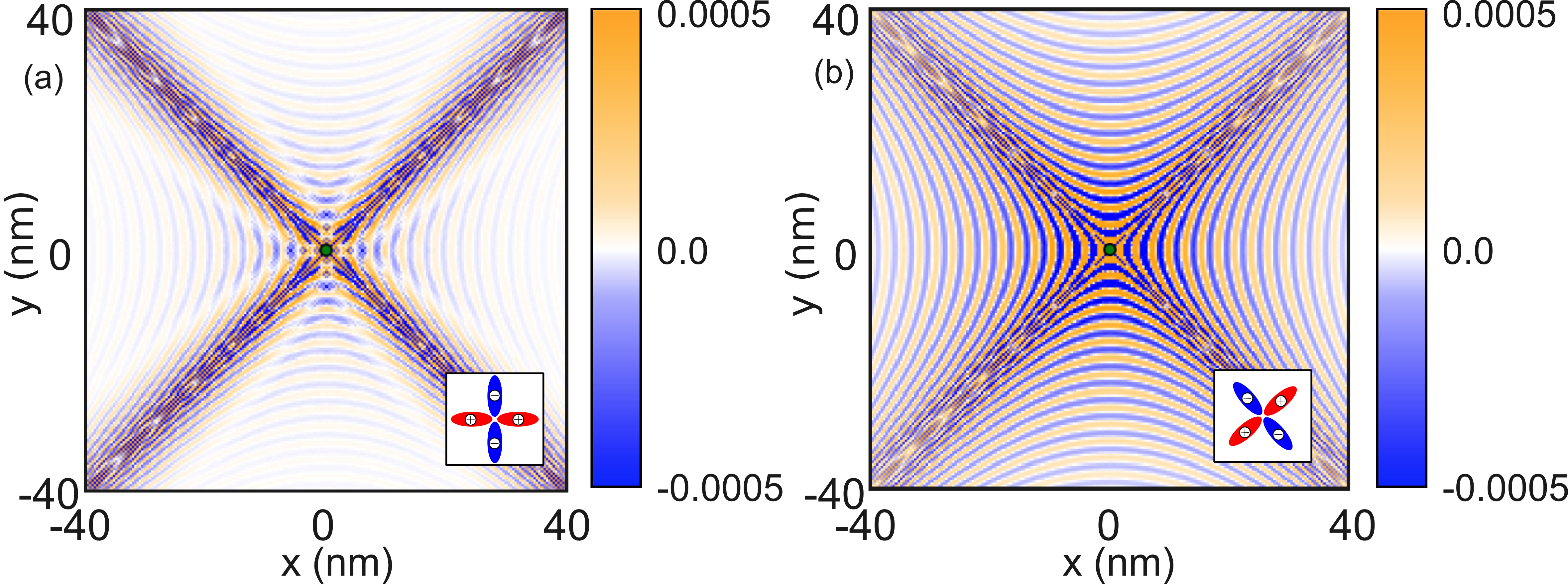} 
	\caption{
		Van Hove-selective Friedel oscillations in the presence of superconducting gap functions with different angular dependences: (a) $d_{x^2-y^2}$, whose maxima occur at the vH points, and (b) $d_{xy}$, whose maxima lie away from the vH points. The $d_{x^2-y^2}$ gap is maximal at the $X$ and $Y$ points in $k$ space, resulting in Friedel oscillation patterns taken at low voltage bias being suppressed along the real-space $x$ and $y$ axes. Conversely, the $d_{xy}$ gap is maximal along the $k$-space diagonal directions, suppressing Friedel oscillations along the corresponding real-space diagonals and leaving comparatively stronger oscillations along the $x$ and $y$ axes.
		}
	\label{fig:Friedel_SC_nonSC0}
	\vspace{-2mm}
\end{figure}

Furthermore, as shown below, vH points $X$ and $Y$ are responsible for light-cone Friedel oscillations in different quadrants, the East-West for $X$ and North-South for $Y$. This property provides a convenient diagnostic of $\Delta_X$ and $\Delta_{Y}$, which can be determined separately and independently. Similarly, for the bandstructures with three vH points, such as those common for bandstructures with hexagonal lattice symmetry (see, e.g., Ref.\cite{Nandkishore2012}), light-cone Friedel oscillations from different vH points occur in different $60^\circ$ sectors, allowing separate diagnostic of the gap function at these vH points. 

This spatial separation gives the STM measurement unique diagnostic power. 
A conventional local tunneling spectrum receives contributions from the entire Fermi surface, so a low-bias suppression alone cannot determine whether the superconducting gap is concentrated near a vH point or elsewhere in momentum space. Here, in contrast, quasiparticles originating near different vH points propagate into distinct, non-overlapping real-space sectors. Comparing the fringe envelope in each sector between the normal and superconducting states therefore provides a momentum-selective measurement without requiring momentum-resolved spectroscopy: suppression in a given sector directly signals a gap for the corresponding vH quasiparticles.

Experimentally, the proposed protocol is therefore as follows. First, a low-bias $dI/dV$ map is acquired around an isolated impurity in the normal state, either above $T_c$ or with superconductivity suppressed by a magnetic field. The resulting light-cone fringe patterns form the real-space sectors associated with individual vH points, giving a normal-state reference for the oscillation period and decay length. Repeating measurement in the superconducting state under the same conditions, 
one can determine whether superconductivity selectively suppresses quasiparticles originating from a given vH point. Fitting the decay lengths in the normal and superconducting states then yields the corresponding gap magnitude through the relation derived below. Repeating this analysis for the distinct vH sectors allows the gap magnitudes at symmetry-inequivalent saddle points to be determined independently.
Accessing phases requires methods such as the HAEM~\cite{Hirschfeld2015}, the two-impurity interference~\cite{Ding_hyperbolic_2023,Panigrahi_tomographic_2026,Panigrahi_particle-hole_2026}, or other phase-sensitive probes.

	\begin{figure}[t!]
	\centering
	\includegraphics[width=0.99\linewidth]{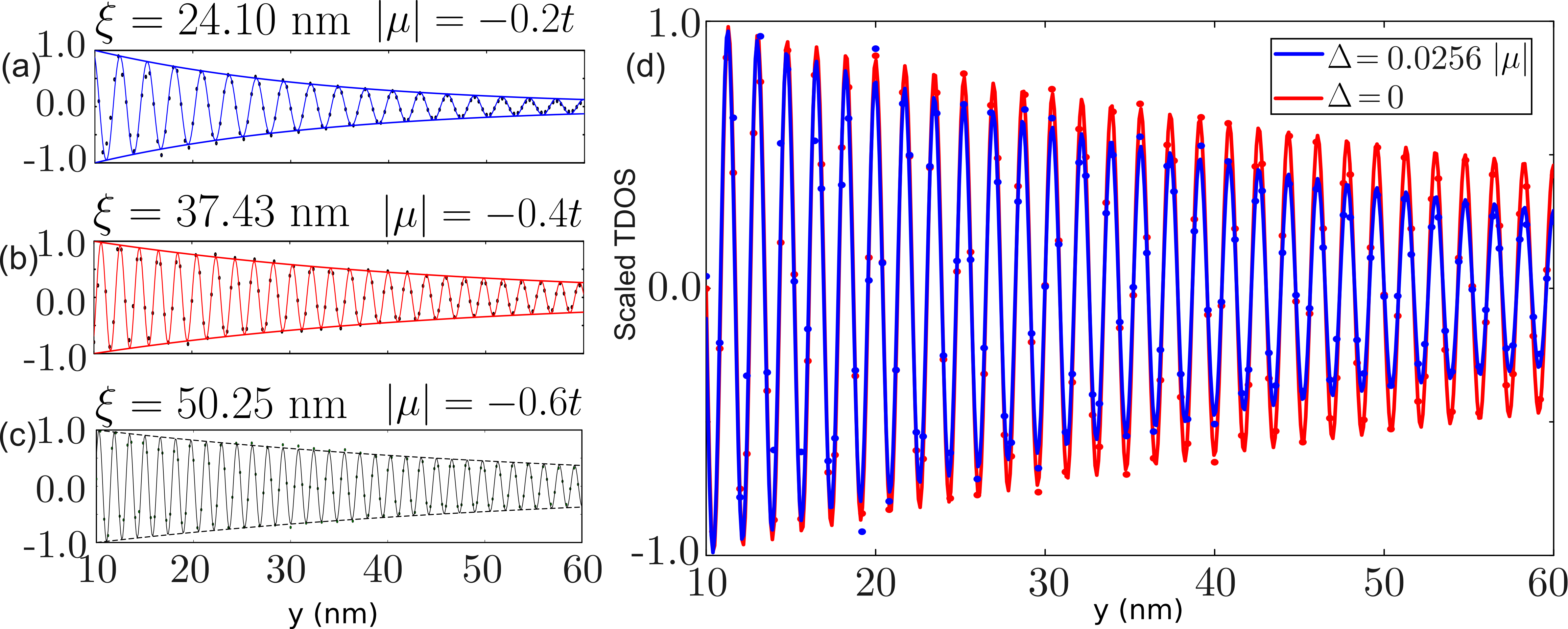} 
	\caption{
	Probing superconductivity by vH-selective Friedel oscillations. Shown is TDOS pattern on a line-cut through the impurity along the $y$ direction. 
	Oscillations fall off exponentially, as $e^{-2y/\xi}$, with the best-fit lengthscale $\xi$ values quoted in (a), (b) and (c) matching closely the prediction from Eq.~\eqref{eq:coherence-length}.
	Since  $\xi$ is proportional to the Fermi velocity, detuning away from the vH point (at the same temperature and same gap magnitude) causes slower decay, as demonstrated in (a)-(c). From the exponential fits, we reconstruct the gap magnitude using Eq.~\eqref{eq:two-temp}, and find that the agreement with Eq.~\eqref{eq:coherence-length} is better than $15\%$ even at detunings as large as $\mu = -0.6 t$, and the disagreement is less than a percent for $\mu = -0.2 t$. 
	Panel (d) illustrates that the superconductor has smaller coherence length than the normal metal at a comparable temperature, as expected from Eq.~\eqref{eq:coherence-length}. The magnitude of the gap is estimated using Eq.\eqref{eq:two-temp}, to a high degree of accuracy.
	}
	\label{fig:Friedel_SC_nonSC}
	\vspace{-2mm}
\end{figure}

Beyond probing superconductivity, the vH-selective Friedel oscillations may provide a real-space probe of anomalous normal-state carrier dynamics near vH points~\cite{Barber2018,Jaoui2022,Wei2024}. Interactions near a vH singularity can produce marginal- or non-Fermi-liquid self-energies and anomalously short quasiparticle lifetimes~\cite{Kapustin2018,KataninKampf2004}. For a hyperbolic saddle, Houghton, Mishchenko and Raikh
~\cite{Houghton2019} found a thermally cut-off equilibrium Friedel response and, separately, a marginal inelastic lifetime near the Lifshitz transition. With thermal and disorder broadening independently calibrated, the temperature- and energy-dependent envelope of the vH-resolved TDOS fringes could test whether anomalous scattering is concentrated near particular vH sectors. The pairing-induced sector suppression studied here then supplies the complementary gap diagnostic, allowing vH-related decoherence and superconducting gap formation to be distinguished in real space. This is an interesting topic for future work. 

The analysis of vH Friedel oscillations in a superconductor can be carried out using the BdG Hamiltonian describing quasiparticles near the Fermi level,
	\begin{equation}\label{eq:BdG-Hamiltonian}
H_{\mathrm{BdG}}(\vec{k})
	=
	\begin{pmatrix}
		E_{\vec{k}}-\mu & \Delta_{\vec{k}}\\
		\Delta_{\vec{k}}^* & \mu-E_{-\vec{k}}
	\end{pmatrix}.
	\end{equation}
	The corresponding momentum space Green's function for the superconductor is,
	\begin{equation}
		G(\vec{k}, i\omega_m) = \frac{1}{i\omega_m - (E_{\vec k}(\vec k)-\mu)\sigma_z - \Delta_k \sigma_x}
	\end{equation}
	A much-studied superconductivity problem with pairing enhanced by vH points is defined on a 2D square lattice, Eq.\eqref{eq:square_lattice_bandstructure}. 
	The saddle points occur at $X=(\pm \pi/a,0)$, and $Y=(0,\pm \pi/a)$, where $a$ is the lattice constant. The system is half-filled at $\mu=0$, electron-doped at $\mu>0$, and hole-doped at $\mu<0$. Superconducting gap function centered at vH points is modeled as
	\begin{equation}
		\Delta^{\rm vH}_\kk = \Delta_0 e^{-(\vec k - \vec k_{\rm vH})^2/k_0^2}. 
	\end{equation}
		This gap function has $s$-wave symmetry if $\Delta_{ X} = \Delta_{ Y}$,  $d_{x^2-y^2}$ symmetry if $\Delta_{ X} = -\Delta_{ Y}$, or no rotational symmetry (nematic superconducting order) if $\Delta_{ X} \ne \Delta_{ Y}$. The results are found to be not particularly sensitive to the gap function peak width parameter $k_0$.
	
	Near the vH point $Y$, we can write $k_x = q_x$ and $k_y = \pi + q_y$. Expanding to quadratic order, we obtain the effective dispersion 
\begin{equation} \label{eq:saddle_dispersion_1}
	E_{(0,\pi)}(\vec{q}) \approx t a^2 (q_x^2 - q_y^2) - \mu.
\end{equation}
	 Similarly, for the other van Hove point ($X$), the effective dispersion is \begin{equation}  \label{eq:saddle_dispersion_2}
	 	E_{(\pi,0)}(\vec{q}) \approx t a^2 (q_y^2 - q_x^2) - \mu.
\end{equation}
	 	 Thus, the two saddle points have hyperbolic dispersions with exchanged principal directions.
	
	This simple fact governs the quasiparticle dynamics in real-space. For an ordinary convex Fermi surface, a point source can emit semiclassical quasiparticle trajectories in an approximately isotropic manner. In comparison, near the $Y$ van Hove point, the low-energy quasiparticles predominantly move along $y$ axis and nearby directions confined to $90^\circ$-quadrants. At other spatial directions the semiclassical trajectories are forbidden. The boundaries between the two regions occur at the lines $y = \pm x$. The allowed sector is $|y|>|x|$ (`time-like' separated region) where the propagator oscillates. Outside this (for $|x|>|y|$, i.e., the `space-like' separated region) the propagator is suppressed exponentially. For the $X$ van Hove point, 
	the roles of $x$ and $y$ are interchanged: semiclassical trajectories are confined to $90^\circ$-quadrants around the $x$ axis, and are absent outside these quadrants. This is the origin of the light cone effect.
	
	Within the hyperbolic saddle approximation for the bandstructure, the real-space Green's function due to a van Hove point is
	\begin{equation}\label{eq:G_BdG_Ypoint}
		G^{(0,\pi)}(\vec r) = \frac{i\omega_n + \Delta \sigma_x}{\Omega_n} F_{0}(\vec r) + \sigma_z F_{1}(\vec r)
	\end{equation}
		where $\Omega_n = \sqrt{\omega_n^2 + \Delta^2}$, and $F_0(\vec r)$ and $F_1(\vec r)$ are, respectively, the imaginary and real parts of the modified Bessel function of the second kind
		\begin{equation}
		K_0\left(\sqrt{-\frac{|\mu|+i\Omega_n}{ta^2}}\sqrt{y^2-x^2}\right)
		\end{equation} 
		 (sometimes also called the Macdonald function).
Similarly, for vH point $X=(\pi,0)$
	\begin{equation}
		G^{(\pi,0)}(\vec r) = \frac{i\omega_n - \Delta \sigma_x}{\Omega_n} F_{0}(\vec r) + \sigma_z F_{1}(\vec r)
	\end{equation}	
	For the $X$ saddle point, the argument in $F_0$ and $F_{1}$ will be $(x^2-y^2)^{1/2}$ instead.
	
	Near the $Y$ point, the propagators vanish exponentially in the regions $|x| > |y|$, a behavior similar to Klein-Gordon relativistic scalar field propagator outside its light cone, and they sinusoidally oscillate in the regions $|x| < |y|$. This can be understood from the fact that near the van Hove point $Y = (0,\pi)$, the semiclassical velocity aligns with the $\pm y$ axis, and the quasiparticle cannot propagate along the $x$ axis.
	In a similar manner, near the $X$ point, the propagators vanish exponentially in the regions $|x| < |y|$ and oscillate in the regions $|x| > |y|$. 
	
	Recently, hyperbolic spatial patterns were found for the LDOS of Yu-Shiba-Rushinov (YSR) states in a vH superconductor on a square lattice~\cite{Uldemolins_quasiparticle_2022}, whereas YSR states for a polygon-shaped Fermi surface were found to produce highly anisotropic beams in LDOS~\cite{Ortuzar_yu-shiba-rusinov_2022}.
	
			To describe Friedel oscillations near a localized impurity we consider the real-space tunneling density of states (TDOS) $dI/dV$. For conventional (circular) Fermi surfaces, the impurity-modified TDOS features the familiar form of Friedel oscillations consisting of concentric rings.
	In comparison, the van Hove singularity produces hyperbolic fringes in the Friedel oscillations near an impurity (see Fig.~\ref{fig:vh}), of the form expressed in Eq.~\eqref{eq:light-cone-behavior}.
	
	
The origin of the light-cone Friedel fringes and their spatial structure, can be illustrated by considering impurity-induced density oscillations in the normal state, \(\Delta=0\), near a weak point-like impurity \(U(\vec r)=u_0\delta(\vec r)\). To first order in \(u_0\), the local Green's function acquires the correction
\begin{equation}\label{eq:delta G=G0uG0}
\delta G^R_1(E;\vec r,\vec r)
=u_0\left[G_0^R(E,\vec r)\right]^2.
\end{equation}
For the hyperbolic saddle-point dispersion, the retarded propagator in a semiclassically allowed angular sector is expressed through a Hankel function, as given above, with large-distance asymptotics
\begin{equation}
G_0^R(E,\mathbf r)\propto
\frac{e^{\pm i(q_Es-\pi/4)}}{\sqrt{q_Es}},
\qquad
q_E=\sqrt{\frac{|E+\mu|}{ta^2}}.
\end{equation}
Here \(s=\sqrt{|x^2-y^2|}\) is the Minkowski distance introduced above, restricted to the allowed sector. Integrating the impurity-induced spectral density over occupied energies gives, at zero temperature:
\begin{equation}
\delta n_{\rm osc}(R,\theta)
\simeq
\frac{u_0}{8\pi^2 ta^2 R^2|\cos2\theta|}
\sin\left(2k_F R\sqrt{|\cos2\theta|}\right),
\end{equation}
where \(k_F=\sqrt{|\mu|/ta^2}\), \(x=R\cos\theta\), and \(y=R\sin\theta\).  The quantity $s=R\sqrt{|\cos2\theta|}$ is nothing but the Minkowski distance introduced above. This result applies for \(k_Fs\gg1\) in the oscillatory sectors of a single saddle-point band, and vanishes in the complementary sectors. The constant-phase contours are hyperbolas, while the amplitude is strongly enhanced toward the light-cone boundaries \(x=\pm y\). The apparent divergence near these boundaries signals the breakdown of the large-distance asymptotic expansion, a behavior that can be cured by reintroducing the full propagator.

The result above describes the equilibrium density modulation, obtained by integrating the impurity-induced local density of states over occupied energies. In contrast, STM probes the energy-resolved tunneling density of states (TDOS), which at low temperatures is proportional to the differential conductance ($dI/dV$) at the corresponding bias voltage. This distinction affects the spatial decay and phase of the oscillations: at zero temperature, the equilibrium density oscillations decay as $1/s^2$, whereas the TDOS oscillations at fixed energy decay as $1/s$. Both, however, exhibit the same characteristic hyperbolic geometry. We now calculate the TDOS in the superconducting state using the BdG Green's functions and the impurity correction in Eq.\eqref{eq:delta G=G0uG0}.
	

Next, we calculate tunneling current in the superconducting state, $\Delta\ne0$.	Using the bare BdG Green's function given above, Eq.\eqref{eq:G_BdG_Ypoint}, the impurity-modified Green's function can be computed, and related to TDOS as
	\begin{equation}
		\label{eq:dIdV}
		\left . \frac{dI}{dV}\right|_{V=0} = 4 e^2 t_{\rm STM}^2 {\text{Re}}\,{\text{Tr}} \left[ T\sum_{\omega_n > 0} \left. \frac{d\delta G_1}{dz} \right|_{z=i\omega_n}\right]
	\end{equation}
	(see also Ref.\cite{Panigrahi_tomographic_2026}). 
	Here we consider point-like impurities which scatter quasiparticles as delta-function potentials, $U(\vec r) = u_0 \delta(\vec r - \vec r_i)$. In momentum space, each impurity generates a constant interaction that scatters between any two momenta with equal rates.
	
	At half filling, the TDOS signal is found to be concentrated only on the light-cone boundaries. However, slightly away from half-filling, 
	Friedel oscillations take the light-cone form of sharp, hyperbolic fringes. Moreover, since $k_F \propto \sqrt{\mu}$, the distance between consecutive fringes 
	is $\lambda_F/2$, which scales as $1/\sqrt{\mu}$.
	
	This is illustrated in Fig.~\ref{fig:vh}, 
	obtained by a numerical calculation 
	using the lattice Green's function of the normal metal phase in a square lattice band. As the hole doping is increased, eventually the Fermi surface becomes nearly circular, 
	described by a conventional pattern of concentric rings~\cite{Bena2016}.
	
		\begin{figure}
	\includegraphics[width=0.99\linewidth]{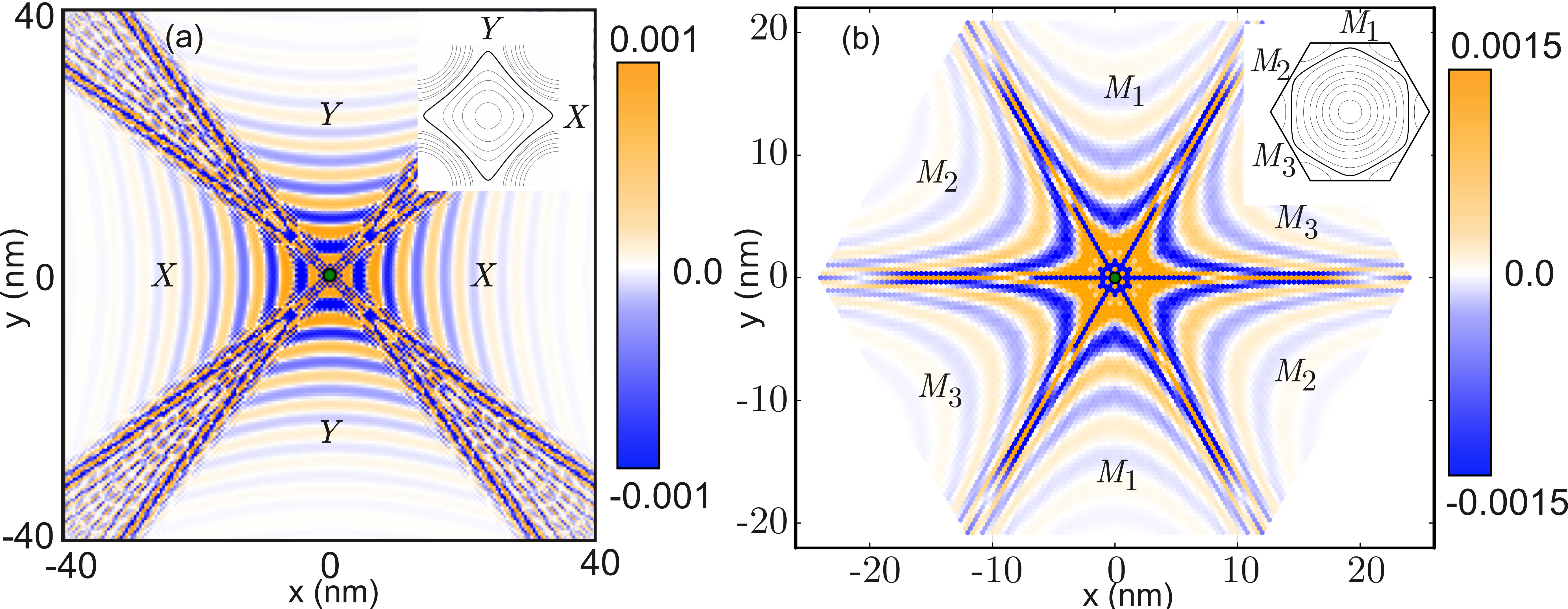} 
	\caption{
Van Hove–selective Friedel oscillations arising from van Hove (vH) points in two band structures: (a) a square-lattice $t-t'$ hopping model with $t'=-0.2t$, and (b) a triangular-lattice nearest-neighbor hopping model. The vH selectivity is evident in both cases: each vH point generates hyperbolic Friedel fringes confined to a distinct pair of real-space sectors [$X$ and $Y$ in (a); $M_1$, $M_2$, and $M_3$ in (b)]. The light cone opening angle is smaller than $90^\circ$ due to nonzero $t'$. In each case, the hyperbola asymptotes define the sector boundaries, demonstrating that this behavior is generic rather than tied to a particular band-structure model.
}
	\label{fig:Friedel_nonSC_different_lattice}
\end{figure}
	
	
	The hyperbolic Friedel fringes generated by vH quasiparticles provide a natural real-space probe of vH-assisted superconductivity. If pairing opens a strong gap near a vH point, the low-bias, low-$T$ quasiparticle density of states is depleted and the corresponding fringes within its light-cone sector are strongly suppressed.

This is demonstrated in Fig.~\ref{fig:Friedel_SC_nonSC}, where the Friedel oscillation near an impurity is compared for $\Delta =0$, and $\Delta=0.0256 |\mu|$, at a fixed temperature. If the van Hove points host a hotspot for the gap function, then such a suppression of TDOS at certain directions will be observed as the system enters the superconducting phase from the normal metal phase. Such a comparison can help to experimentally determining whether the superconducting gap function is localized near the van Hove points, or away from them. To the contrary, if the gap function is primarily localized away from the van Hove points, then such a strong suppression of TDOS will not be observed in the superconducting state.


To assess feasibility of using vH Friedel oscilations for determining $\Delta$, we note that small $k_F$ and $v_F$ values near vH points have two crucial effects. First, as discussed above, the oscillation periodicity, tunable by doping to vH point, becomes significantly larger than the atomic scale. Second, the exponential decay due to finite coherence length~\cite{Fetter1965,Machida2003,Stociek2022} 
occurs at the lengthscales orders of magnitude shorter than those in superconducting metals, bringing it in the STM-accessible range---see Eq.\eqref{eq:coherence-length}. This 
gives a recipe for determining $\Delta$.  
Comparing the measured decay lengths $\xi(T,\Delta)$ in the normal state at temperature $T_2$ and in the superconducting state at $T_1$ yields the gap magnitude:
\begin{equation}\label{eq:two-temp}
\Delta^2(T_1) = (\pi k_B T_2)^2 \frac{\xi^2(T_2,\Delta=0)}{\xi^2(T_1,\Delta(T_1))} - (\pi k_B T_1)^2.
\end{equation}
Typically, the normal-state reference is acquired at $T_2>T_1$. If superconductivity is suppressed by a magnetic field, it can instead be measured at the same temperature, $T_2=T_1$. This procedure yields the gap magnitude but not the relative phases of $\Delta$ at different vH points. 
	
Lastly, we show that the light-cone Friedel oscillations are not a peculiarity of the nearest-neighbor square-lattice model. Figure~~\ref{fig:Friedel_nonSC_different_lattice} compares a square-lattice $t-t'$ band [Fig.~~\ref{fig:Friedel_nonSC_different_lattice}(a)] with a nearest-neighbor triangular-lattice band [Fig.~~\ref{fig:Friedel_nonSC_different_lattice}(b)], whose global Fermi-surface geometries and arrangements of saddle points are qualitatively different. In both cases, however, the response separates into real-space sectors associated with individual vH points. The square lattice has two inequivalent saddle-point directions, producing pairs of less-than $90^\circ$ sectors associated with the $X$ and $Y$ points, whereas the three symmetry-related $M$ points of the triangular lattice produce pairs of $60^\circ$ sectors. As a further test of generality, we checked that the light-cone character of Friedel oscillations in the square lattice remains unchanged for a more general $t-t'-t''$ band model. 

The general character of the light-cone behavior and its independence of the band symmetry follows from the generic local structure of a two-dimensional saddle point. After choosing its principal axes, the dispersion has opposite curvatures along the two directions. This hyperbolic geometry confines low-energy quasiparticle propagation to a pair of real-space sectors, while propagation in the complementary directions is exponentially suppressed. Crystal symmetry determines how the sectors associated with the different vH points are oriented and tiled in real space, but without constraining the sectors' selectivity. Thus, even when the global band structure differs substantially from the idealized square-lattice case, an STM map can identify the spatial sectors that selectively probe quasiparticles near each vH point.
		
In conclusion, near an impurity, each van Hove point maps onto its own real-space “light-cone” sector, allowing STM around a single impurity to measure whether—and how strongly—that saddle-point quasiparticle sector is gapped by superconducting order. Namely, quasiparticles near vH points generate distinctive real-space signatures in the tunneling density of states near impurities. The unusual momentum-space geometry of the van Hove points is reflected in the strongly anisotropic real-space propagation in both the normal metal and the superconductor. The hyperbolic dispersion near the van Hove points restricts the possible quasiparticle velocity direction, which effectively divides real space into sectors where propagation is allowed, resulting in oscillating Green's function, and forbidden sectors where propagators exponentially decay over a short length scales. The resulting Friedel oscillations of TDOS near an impurity display hyperbolic patterns rather than conventional maps consisting of concentric circles. Strong superconducting pairing near the van Hove points suppresses the hyperbolic fringes in Friedel oscillations as no quasiparticles modes near van Hove points are available at a low-energy. In comparison, when the gap function is concentrated away from the van Hove regions, the characteristic hyperbolas in Friedel oscillations remain very similar to its normal metal counterpart. In this manner, the single-impurity interference can be utilized to experimentally distinguish whether the gap function is strongly localized at van Hove points, or away from them.

We acknowledge useful discussions with Andrey Chubukov, Victor Galitski, Leonid Glazman, Peter Hirschfeld, Patrick Lee and Boris Spivak.
	

	\end{document}